# A highly specific gold nanoprobe for live-cell single-molecule imaging

Cécile Leduc[†], Satyabrata Si[†], Jérémie Gautier[‡], Martinho Soto-Ribeiro[#], Bernhard Wehrle-Haller[#], Alexis Gautreau[‡], Grégory Giannone[§], Laurent Cognet[†] and Brahim Lounis*[†]

[†] *Univ Bordeaux, LP2N UMR 5298, Institut d'Optique & CNRS, F- 33405 Talence, France*
[‡] *CNRS, Laboratoire d'Enzymologie et Biochimie Structurales UPR3082, F-91198 Gif-sur-Yvette France*
[#] *University of Geneva, Department of Cell Physiology and Metabolism, Centre Médical Universitaire, Geneva, Switzerland*
[§] *Univ Bordeaux, IINS UMR 5297, CNRS, F-33000 Bordeaux, France*

*To whom correspondence should be addressed. E-mail: blounis@u-bordeaux1.fr

**Abstract**

Single molecule tracking in live cells is the ultimate tool to study subcellular protein dynamics, but it is often limited by the probe size and photostability. Due to these issues, long-term tracking of proteins in confined and crowded environments, such as intracellular spaces, remains challenging. We have developed a novel optical probe consisting of 5-nm gold nanoparticles functionalized with a small fragment of camelid antibodies that recognize widely used GFPs with a very high affinity, which we call GFP-nanobodies. These small gold nanoparticles can be detected and tracked using photothermal imaging for arbitrarily long periods of time. Surface and intracellular GFP-proteins were effectively labeled even in very crowded environments such as adhesion sites and cytoskeletal structures both *in vitro* and in live cell cultures. These nanobody-coated gold nanoparticles are probes with unparalleled capabilities; small size, perfect photostability, high specificity, and versatility afforded by combination with the vast existing library of GFP-tagged proteins.

Over the last decade, it has become essential to work at the individual molecule scale in order to better understand complex cellular processes. Numerous and increasingly sophisticated optical techniques have been developed to explore the dynamics of biomolecules either *in vitro* or in living systems. Due to its high sensitivity and non-invasiveness, fluorescence microscopy is widely used to detect molecules evolving in biological environments. Single molecule detection requires optimized probes and much effort has been made to improve their properties [1]. In general, the ideal probe should be: (i) very specific, (ii) as small as possible (so not to interfere with the dynamics and functions of the target biomolecules), (iii) monovalent, (iv) optically stable for long observation times, (iv) able to deliver intense optical signals for precise localization. As this ideal probe has not yet been designed, long-term single-molecule tracking of proteins in confined and crowded environments remains challenging.

Fluorescent molecules have been extensively used for their small size but suffer from poor photostability, especially in physiological environments[2]. Semi-conducting nanocrystals (quantum dots), however, display superior photostability and can be imaged for minutes in live cells[3-5]. Nevertheless, due to their relatively large diameter, functionalized quantum dots are not suitable for tracking biomolecules in confined cellular environments such as adhesion sites and synapses [2]. An alternative to luminescence relies on the probe absorption properties. Gold nanoparticles of 5 nm diameter are appealing because they present large absorption cross-sections and deliver intense, stable optical signals in photothermal microscopy, even in cellular environments[6].

To fully benefit from the reduced size of these nanoparticles, the ligand used to bind the particle to the target biomolecule should also be small. Commonly used antibodies have typical sizes of 12 nm and thus significantly increase the overall probe size. Moreover, highly specific primary antibodies are not always available, and when they are, particle coupling must be performed for each protein of interest. To avoid multiple couplings a secondary antibody could be used at the expense of an increase in final probe size. Common antibodies are also divalent and their use raises issues about the probe's valence state. In this context, the elaboration of general strategies to obtain small, highly specific, monovalent probes that allow for the study of the largest number of proteins is an intense field of research [5, 7-9].

In this paper, we designed and characterized a system that most closely approaches the aspects of an ideal probe. We synthesized 5 nm gold nanoparticles functionalized with nanobodies (NB), a small fragment of camelid antibody (2 nm x4 nm) which recognizes widely used GFPs (3 nm x 4 nm) with a very high affinity ($K_d \sim 0.23$ nM) [10, 11]. Our approach allows the production of purified, monovalent nanoparticles. These small, functional gold nanoparticles (NB-Au-NPs) can be detected and tracked for unlimited periods of time in living cells using photothermal heterodyne imaging [12,6,13]. We demonstrate the versatility and targeting efficiency of our NB-Au-NPs by labeling several types of GFP-tagged proteins, both in *in vitro* and cellular systems. We also show that NB-Au-NPs can be used to study proteins in confined structures and intracellular compartments by tracking adhesive proteins and microtubule associated proteins in living cells, respectively.

NB-Au-NPs were produced in two steps, synthesis and functionalization (Figure 1A, see supporting methods). In the first step, functionalized gold nanoparticles were synthesized using borohydride reduction in the presence of a linker agent, MUA, (11-mercaptoundecanoic acid) and two blocking agents in equimolar amounts: CVVVT-ol peptide (T-ol = threoninol) and alkyl PEG (PEGylated alkanethiol HS-(CH2)$_{11}$-EG4-OH) (protocol adapted from[7, 14]). These peptide capped gold nanoparticles are extremely stable, water-soluble and display some chemical properties analogous to proteins, which is responsible for their biocompatibility[7, 15]. The average diameter of the nanoparticles was estimated to be 4.6 +/- 1 nm (SD) from transmission electron microscopy pictures (Figure 1B-C). Because the molecules of the capping shell are very short the hydrodynamic radius of the capped nanoparticles is less than 1 nm larger[16] than the radius obtained from electron microscopy. A photothermal image of isolated particles embedded in a polyvinyl alcohol film is also displayed in Figure 1D. In the second step, the carboxylic function of MUA was activated using EDC/NHS (1-ethyl-3-(3-dimethylaminopropyl) carbodiimide hydrochloride/N-Hydroxysuccinimide) allowing coupling to recombinant nanobodies. We used nanobodies bearing a his-tag purified from bacteria with Ni-NTA columns. We then employed two methods to obtain monovalent probes. In method one, the proportion of MUA was reduced with respect to that of the blocking agents (down to 0.4%, Supporting Figure 1) to achieve at most one MUA per particle when an excess of nanobodies was introduced for the coupling. In method two, a low concentration of nanobodies was introduced into an excess of particles bearing multiple MUA linkers. We favored the second method, which requires a smaller amount of nanobodies, because non-functionalized Au-NPs could be separated from the NB-Au-NPs by using Co-NTA resin able to reversibly bind his-tag nanobodies. This is demonstrated in Figure 1E by the shift of nanobody-coated Au-NP band as compared to the control Au-NP in agarose gel electrophoresis. It is also seen that a coating with 20% MUA produces negatively charged particles (line (a) and (f) to (h)), while with 1% and 0.4% MUA, positively charged NB-Au-NPs are obtained and cannot be used for cell electroporation (see below). Of note, NB-Au-NPs bands were relatively broad due to the size dispersion of the nanoparticles (Figure 1C) leading to a relatively large heterogeneity in the number of MUA per particle.

NB-Au-NPs were first tested on a well-characterized in vitro assay consisting of purified molecular motors and microtubules in a reconstituted system[17,18]. GFP-tagged kinesin-1 were incubated with an excess of NB-Au-NPs for 5 min before injection in a flow chamber where taxol-stabilized microtubules were previously fixed (Supporting Figure 2)[19]. Without ATP in solution, kinesin motors labeled with NB-Au-NPs strongly bound microtubules without any directional motion or unbinding (Figure 2A). We used a high concentration of motors in solution (15 nM, 5 min incubation) to saturate microtubules with motors and then rinsed extensively in the flow chamber to remove unbound motors and probes. A homemade microscopy setup was used to record photothermal and fluorescence images of the sample (see supporting information). Figure 2B shows a photothermal image of microtubules covered by kinesin-GFP motors labeled with NB-Au-NPs and the corresponding fluorescence image. The GFP fluorescence signal and NB-Au-NPs photothermal signal colocalize within a camera pixel size (~250nm at the sample plane), a signature of a specific labeling of the motors by the new probe. Regions with higher photothermal signals correspond to overlapping microtubules (arrow in Figure 2B). More quantitatively, the signal in these regions is, in average, twice the signal of isolated microtubules (Figure 2C). This illustrates the possibility to perform stoichiometric measurements using NB-Au-NP probes detected by photothermal methods[12, 20]. As a control, we verified that no photothermal signal was observed when microtubules were incubated with non-functionalized Au-NPs or with NB-Au-NPs without kinesin motors (supporting Figure2).

To demonstrate that NB-Au-NPs binding specificity is conserved in more complex environments, we tested the particles in fixed cells expressing GFP-tagged proteins in various sub-cellular compartments. We used either COS-7 cells or Mouse Embryonic Fibroblasts (MEFs) platted on fibronectin-coated glass coverslips. The cells were transfected with GFP-tagged proteins enriched in cellular adhesion sites (VASP-GFP), in the actin cytoskeleton (alpha-actinin-GFP), or on microtubules (EB3-GFP) (Figure 3). After fixation, and permeabilization, followed by incubation with a 10 nM solution of NB-Au-NP and extensive rinsing steps, cells were studied by epi-fluorescence and photothermal imaging. For all protein constructs, photothermal images displayed perfect colocalization with the GFP fluorescence images, demonstrating that the probe specificity is maintained in complex biological environments. We verified that non-transfected cells displayed very few non-specific nanoparticle binding (Supporting Figure3). Notably, these results show that NB-Au-NP can access GFP-proteins even in crowded regions of the cells such as adhesion sites or actin rich structures like the lamellipodium.

Next, we showed that NB-Au-NPs could label membrane proteins in live cells, allowing single molecule tracking even in highly confined structures. As a model, we focused on mature adhesion sites (focal adhesions, FAs) which are crowded macromolecular platforms where integrins mediate cell-extracellular matrix adhesion[21]. β3-integrins bearing an extracellular GFP tag were expressed in MEFs and found to be concentrated in FAs (Figure 4A epi-fluorescence). MEFs were incubated with NB-Au-NPs for 10 min and then rinsed extensively. Photothermal images of the same region as in Figure 4A showed that the NB-Au-NPs were also mainly localized in the FAs (delimited regions of the epi-fluorescence image), strong evidence that the probe can enter FAs with minimal steric hindrance. In contrast, his-tagged β3-integrins labeled with streptavidin coated quantum dots via a biotinylated NTA[22] could not enter FAs (Figure 4B). Since the space between the membrane and the glass coverslip is close to 30 nm in FAs[21], this observation suggests that the size of the quantum dot probe prevents its access to the FA interior.
In order to follow the movement of individual GFP-β3-integrins we reduced the labeling density of NB-Au-NPs on MEFs. Using the triangulation method introduced by Lasne et al[6], long trajectories (>25 s) of single integrins were recorded at video rate (Figure 5). In Figure 5A, we overlaid two typical trajectories of individual integrins on the fluorescence image of GFP-β3-integrin. The two trajectories illustrate the diffusion properties of integrins found inside and outside FAs. The photothermal signals displayed a constant amplitude and were similar to those shown in Figure 1D obtained from single gold particle with size distribution of Figure 1C, confirming that these trajectories corresponded to individual integrins (Figure 5 B and E inset). We also extracted the mean-square-displacements and instantaneous diffusion coefficients (with a moving window of ~ 1s, see supporting information) of the probes. Inside FAs, probes displayed either immobilization or slow diffusion, whereas outside they displayed free diffusion (Figure 5B-E, supporting Figure 4). These behaviors are similar to those observed using single molecule fluorescence tracking with smaller probes[23], which showed that immobilization events observed inside FAs are the consequence of direct integrin binding to the extracellular matrix (fibronectin) and/or to intracellular talins. This confirms that NB-Au-NPs did not alter integrin dynamics after labeling nor did they impede their activation. In addition, combining photothermal imaging and the perfect photostability of our probes, we now are able to investigate the dynamics of single integrins on arbitrarily long time scales. In the specific case of integrins, we can study cycles between immobilizations and free diffusive movements even during the same trajectory. This allows to probe the dynamics of protein interactions, which control membrane receptor diffusion.

Finally, we addressed the possibility of targeting and tracking intracellular proteins with NB-Au-NPs in living cells. Negatively charged NB-Au-NPs and 5 µg DNA coding for EB3-GFP, a microtubule end-binding protein, were efficiently internalized in COS-7 cells by electroporation. Figure 6A shows fluorescence and photothermal images of a cell 24h after electroporation. Demonstrating specific intracellular targeting (see supporting Figure 5 for the control), NB-Au-NPs appeared mainly located either along the microtubule lattice or ends, where EB3-GFP proteins are accumulated[24] (red arrows in Figure 6A). We further tracked the movement of bound nanoparticles. The trajectory shown in Figure 6B-D represents the movement of a single EB3-GFP protein: it is clearly directional, points towards the cell periphery in the prolongation of a microtubule, and shows an average velocity (~ 0.09 µm/s), similar to the microtubule growth rate (~0.12 µm/s for COS-7 at room temperature, supporting movie 1). This demonstrates that NB-Au-NPs are appropriate probes to target and track individual intracellular proteins in living cells.

In this paper, we presented the development of a new functional optical probe, which bears all the qualities to perform non-invasive long-term single particle tracking in different biological environments. It consists of 5 nm gold nanoparticles coated with GFP-nanobodies, the NB-Au-NPs. We showed that their charge and valence can be tuned by varying the composition of the peptide shelves, and that they can be purified using the his-tag present on the nanobodies.

These NB-Au-NPs were successfully used in *in vitro* and live cell experiments, for both surface and intracellular labeling of various GFP-proteins in very crowded environments (adhesion sites, actin networks etc.). They should allow high density labeling of many target proteins in live or fixed cells, which will provide more complete information about the distribution. We believe that NB-Au-NPs are unparalleled probes owing to their small size, perfect photostability, high specificity, and versatility. Possible applications could involve the targeting of nuclear proteins or concern correlative studies between optical and electron microscopy since small gold nanoparticles provide high contrasts for both modalities.

**Acknowledgement:**

We would like to thank Marie-Hélène Delville for the use of the TEM microscope, Matthieu Sainlos for the Cobalt resin used to purify the NB-Au-NPs, C. Hoongenrad for the plasmid of EB3-GFP, S. Diez for the plasmid of the kinesin-GFP (rk540-GFP), B. Tessier and O. Rossier for experimental help, and J. Shaver for reading the manuscript. We acknowledge financial support from the Région Aquitaine, Institut universitaire de France, Agence Nationale de la Recherche, and the European Research Council.

# References

1. Giepmans, B. N.; Adams, S. R.; Ellisman, M. H.; Tsien, R. Y. *Science* **2006,** 312, (5771), 217-24.
2. Groc, L.; Lafourcade, M.; Heine, M.; Renner, M.; Racine, V.; Sibarita, J. B.; Lounis, B.; Choquet, D.; Cognet, L. *J Neurosci* **2007,** 27, (46), 12433-7.
3. Dahan, M.; Levi, S.; Luccardini, C.; Rostaing, P.; Riveau, B.; Triller, A. *Science* **2003,** 302, (5644), 442-5.
4. Heine, M.; Groc, L.; Frischknecht, R.; Beique, J. C.; Lounis, B.; Rumbaugh, G.; Huganir, R. L.; Cognet, L.; Choquet, D. *Science* **2008,** 320, (5873), 201-5.
5. Pinaud, F.; Clarke, S.; Sittner, A.; Dahan, M. *Nat Methods* **2010,** 7, (4), 275-85.
6. Lasne, D.; Blab, G. A.; Berciaud, S.; Heine, M.; Groc, L.; Choquet, D.; Cognet, L.; Lounis, B. *Biophys J* **2006,** 91, (12), 4598-604.
7. Duchesne, L.; Gentili, D.; Comes-Franchini, M.; Fernig, D. G. *Langmuir* **2008,** 24, (23), 13572-80.
8. Medintz, I. L.; Uyeda, H. T.; Goldman, E. R.; Mattoussi, H. *Nat Mater* **2005,** 4, (6), 435-46.
9. Howarth, M.; Liu, W.; Puthenveetil, S.; Zheng, Y.; Marshall, L. F.; Schmidt, M. M.; Wittrup, K. D.; Bawendi, M. G.; Ting, A. Y. *Nat Methods* **2008,** 5, (5), 397-9.
10. Rothbauer, U.; Zolghadr, K.; Tillib, S.; Nowak, D.; Schermelleh, L.; Gahl, A.; Backmann, N.; Conrath, K.; Muyldermans, S.; Cardoso, M. C.; Leonhardt, H. *Nat Methods* **2006,** 3, (11), 887-9.
11. Ries, J.; Kaplan, C.; Platonova, E.; Eghlidi, H.; Ewers, H. *Nat Methods* **2012**.
12. Berciaud, S.; Cognet, L.; Blab, G. A.; Lounis, B. *Phys Rev Lett* **2004,** 93, (25), 257402.
13. Duchesne, L.; Octeau, V.; Bearon, R. N.; Beckett, A.; Prior, I. A.; Lounis, B.; Fernig, D. G. *PLoS Biol* **2012,** 10, (7), e1001361.
14. Zheng, M.; Huang, X. *J Am Chem Soc* **2004,** 126, (38), 12047-54.
15. Levy, R.; Thanh, N. T.; Doty, R. C.; Hussain, I.; Nichols, R. J.; Schiffrin, D. J.; Brust, M.; Fernig, D. G. *J Am Chem Soc* **2004,** 126, (32), 10076-84.
16. Octeau, V.; Cognet, L.; Duchesne, L.; Lasne, D.; Schaeffer, N.; Fernig, D. G.; Lounis, B. *ACS Nano* **2009,** 3, (2), 345-50.
17. Korten, T.; Nitzsche, B.; Gell, C.; Ruhnow, F.; Leduc, C.; Diez, S. *Methods Mol Biol* **2011,** 783, 121-37.
18. Varga, V.; Leduc, C.; Bormuth, V.; Diez, S.; Howard, J. *Cell* **2009,** 138, (6), 1174-83.
19. Helenius, J.; Brouhard, G.; Kalaidzidis, Y.; Diez, S.; Howard, J. *Nature* **2006,** 441, (7089), 115-9.
20. Blab, G. A.; Cognet, L.; Berciaud, S.; Alexandre, I.; Husar, D.; Remacle, J.; Lounis, B. *Biophys J* **2006,** 90, (1), L13-5.
21. Kanchanawong, P.; Shtengel, G.; Pasapera, A. M.; Ramko, E. B.; Davidson, M. W.; Hess, H. F.; Waterman, C. M. *Nature* **2010,** 468, (7323), 580-4.
22. Reichel, A.; Schaible, D.; Al Furoukh, N.; Cohen, M.; Schreiber, G.; Piehler, J. *Anal Chem* **2007,** 79, (22), 8590-600.
23. Rossier, O.; Octeau, V.; Sibarita, J. B.; Leduc, C.; Tessier, B.; Nair, D.; Gatterdam, V.; Destaing, O.; Albiges-Rizo, C.; Tampe, R.; Cognet, L.; Choquet, D.; Lounis, B.; Giannone, G. *Nat Cell Biol* **2012,** 14, (10), 1057-67.
24. Stepanova, T.; Slemmer, J.; Hoogenraad, C. C.; Lansbergen, G.; Dortland, B.; De Zeeuw, C. I.; Grosveld, F.; van Cappellen, G.; Akhmanova, A.; Galjart, N. *J Neurosci* **2003,** 23, (7), 2655-64.

# Figures

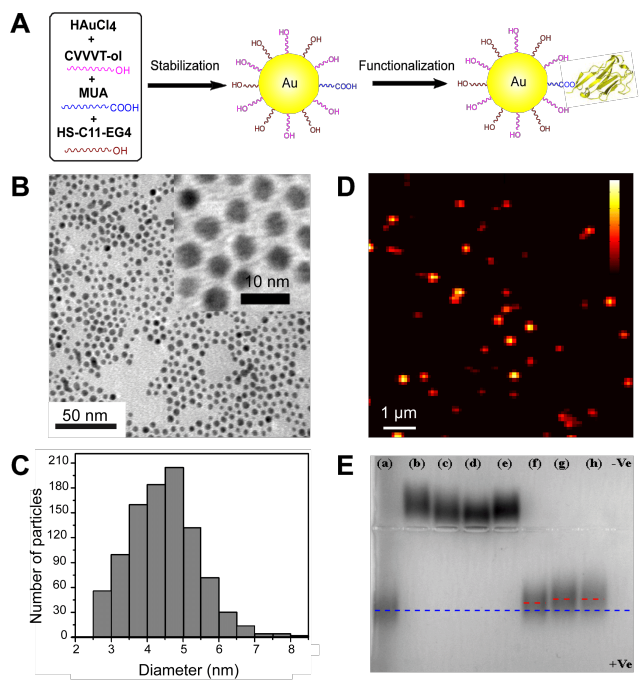

*Figure 1:* Synthesis and characterization of nanobodies-coated gold nanoparticles (NB-Au-NPs). (A) Schematics of the two-steps production: synthesis and functionalization. (B) Transmission Electron microscopy (TEM) image of stabilized Au-NPs. (C) Distribution of stabilized Au-NP diameters, extracted from the TEM images. (D) Photothermal image of individual NB-Au-NPs embedded in thin polyvinyl alcohol (PVA) film. (E) Agarose gel of NB-Au-NP samples with varying proportions of MUA in the nanoparticles shelf and of nanobodies used for the functionalization in step 2: (a) 20% MUA, no nanobodies, (b) 1 % MUA, no nanobodies, (c) 0.4 % MUA, no nanobodies (d) 0.4 % MUA, equimolar nanobodies, (e) 1 % MUA, equimolar nanobodies, (f) 20 % MUA, equimolar nanobodies, (g) 20 % MUA, 2 fold excess nanobodies, and (h) 20 % MUA + 10 fold excess nanobodies.

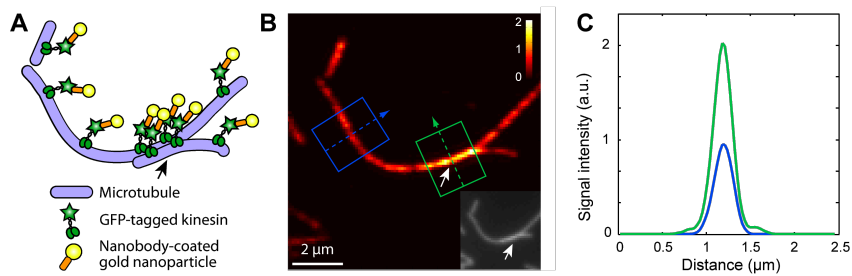

*Figure 2:* *NB-Au-NPs target kinesin-GFPs in vitro. (A) Schematics of the assay: GFP-tagged kinesin motors, attached to stabilized microtubules fixed on a glass coverslip, are labeled with NB-Au-NPs (20%MUA in the shelf, 4x excess of Au-NP with respect to the nanobodies). (B) Photothermal image showing the distribution of NB-Au-NPs on the assay. The photothermal signal exactly matches the distribution of the kinesin-GFP epi-fluorescence signals (inset). White arrows point the overlapping of two microtubules. (C) Average photothermal intensity profile along the blue and green boxes in (B) corresponding to one and two microtubules respectively.*

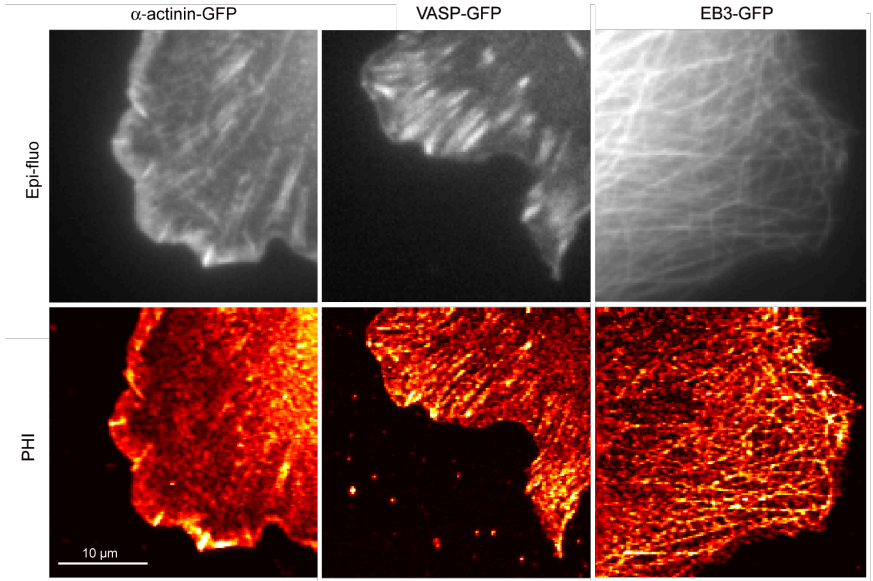

*Figure 3:* NB-Au-NPs label all sorts of GFP-tagged proteins on fixed cells with a very high specificity (NB-Au-NPs used: 20% MUA in the shelf, 4x excess of Au-NP with respect to the nanobodies, see text). Top: epi-fluorescence images of a COS-7 cell expressing α-actinin-GFP, of a MEF expressing VASP-GFP and of a COS-7 expressing EB3-GFP. Bottom: corresponding photothermal images.

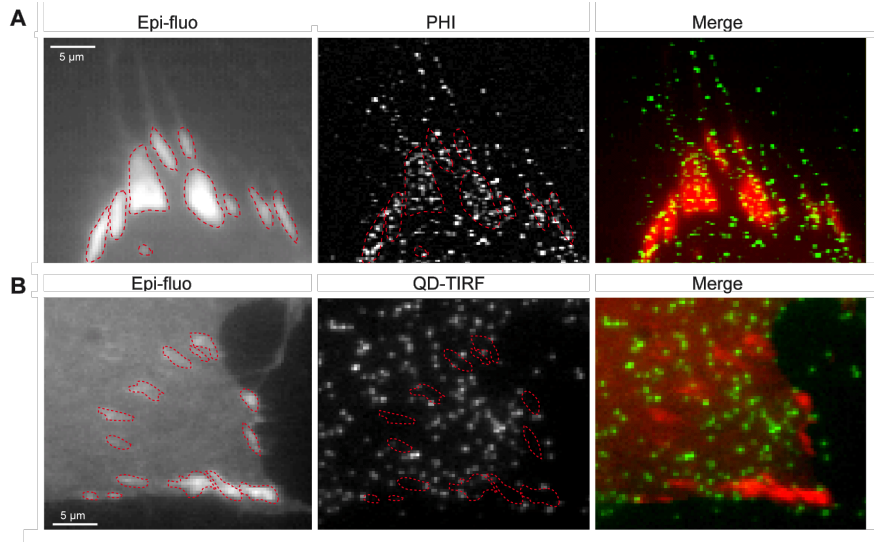

***Figure 4:*** *NB-Au-NPs target β3-integrins in focal adhesion sites whereas quantum dots do not. (A) Live MEF expressing β3-integrins tagged with an extracellular GFP and labeled with NB-Au-NPs. An epi-fluorescence image from GFP-β3-integrins (left) and the corresponding photothermal image (middle) are merged (right) showing efficient NB-Au-NP labeling of GFP-β3-integrins concentrated in FAs (delimited by the red-dashed lines). (B) Live MEF co-expressing VASP-GFP and β3-integrins tagged with an extracellular his-tag and labeled with streptavidin coated quantum dots QD-655 via biotinylated TrisNTA. An epi-fluorescence image of VASP-GFP, used as a FAs reporter, allows the delimitation of FAs contours (left). Corresponding TIRF (total internal reflection fluorescence) image of the quantum dots (middle) and merged image (right) are shown. Quantum dot labeling does not reach β3-integrins located in FAs. NB-Au-NPs were prepared with 20% MUA in the shelf and a 4 fold excess of Au-NP with respect to the nanobodies.*

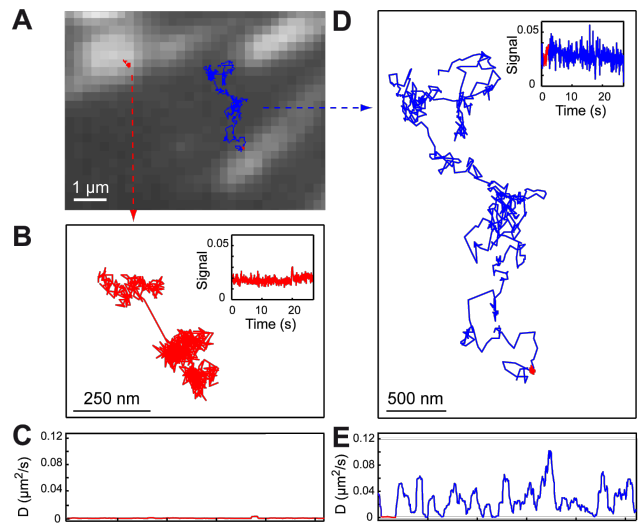

***Figure 5:*** *GFP-β3-Integrins labeled with NB-Au-NPs display a freely diffusive motion outside FAs and cycles of immobilization and slow diffusion motions inside FAs. (A) Two NB-Au-NPs labeled GFP-β3-integrin trajectories are overlaid on FAs localized by GFP-β3-integrin fluorescence. Immobilization sequences and slow movements are displayed in red and free diffusion in blue. (B) x-y trajectory inside a FA. Inset: photothermal signal amplitude vs time. (C) Instantaneous diffusion coefficient vs. time. (D) x-y trajectory outside a FA, and (E) corresponding instantaneous diffusion coefficient vs. time.*

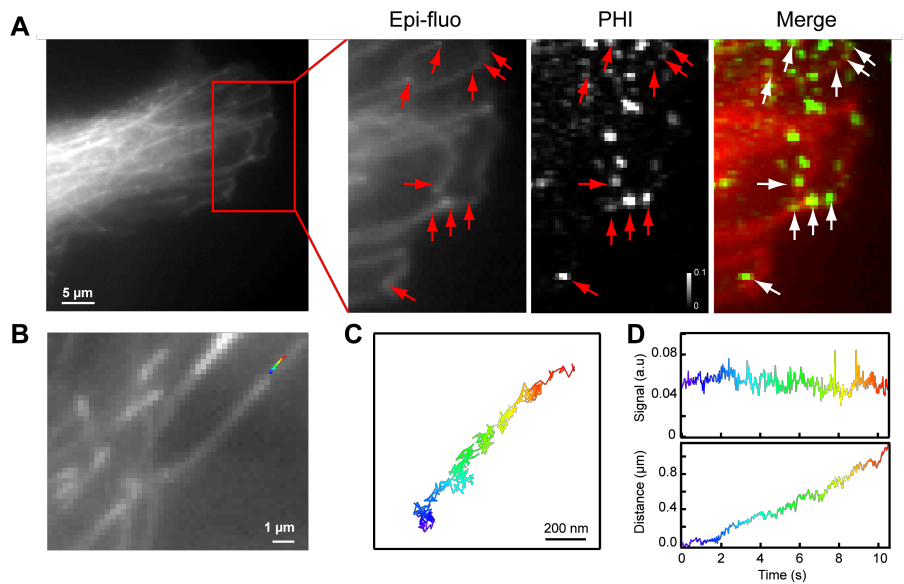

*Figure 6:* Single molecule tracking of EB3-GFP inside living cells using internalized NB-Au-NPs. (A) COS-7 cell expressing EB3-GFP 24h after electroporation. Right: zoom on the cell edge using epi-fluorescence and photothermal imaging together with the merged image. Red arrows show microtubule lattice and ends where NB-Au-NPs are colocalized. (B) EB3-GFP fluorescence image in a COS-7 cell overlaid with the trajectory of an internalized NB-Au-NP, color-coded by the elapsed time (t=0 blue, t=10s red). (C) Zoom on the x-y trajectory. (D) Photothermal signal amplitude and walked distance vs. time of the trajectory displayed in (B) highlighting directed movement. The average velocity of the directed track is ~ 0.09 µm/s. NB-Au-NPs were prepared with 20% MUA in the shelf and a 4 fold excess of Au-NP with respect to the nanobodies.